\documentstyle[psfig]{aa}

\begin{document}

   \thesaurus{03     
              (13.25.2;  
               11.19.1;  
               11.09.1)}  
   \title{BeppoSAX uncovers the hidden Seyfert 1 nucleus in the Seyfert 2 galaxy NGC 2110}

   \author{G.~Malaguti\inst{1}, L.~Bassani\inst{1}, M.~Cappi\inst{1}, A.~Comastri\inst{2},
           G.~Di~Cocco\inst{1}, A.C.~Fabian\inst{3}, G.G.C.~Palumbo\inst{4},
T.~Maccacaro\inst{5}, R.~Maiolino\inst{6}, 
           P.~Blanco\inst{7}, M.~Dadina\inst{8}, D.~Dal~Fiume\inst{1},
F.~Frontera\inst{1},
           M.~Trifoglio\inst{1}
          }

   \offprints{G. Malaguti (malaguti@tesre.bo.cnr.it)}

   \institute{ITESRE/CNR, via Piero Gobetti 101, I-40129 Bologna, Italy \and 
              Osservatorio Astronomico di Bologna, via Ranzani 1, I-40127
Bologna, Italy \and
              Institute of Astronomy, Madingley Road, Cambridge CB3 0HA, UK \and
              Dipartimento di Astronomia, Universit\`a di Bologna, via Ranzani 1,
I-40127 Bologna, Italy \and
              Osservatorio Astronomico di Brera, via Brera 28, 20121, Milano, Italy \and
              Osservatorio Astrofisico di Arcetri, L.E.Fermi 5, 50125, Firenze, Italy \and
              UCSD/CASS, 9500 Gilman Drive, Code 0111, La Jolla, Ca, USA \and
              BeppoSAX SDC, ASI, via Corcolle 19, 00131 Roma Italy 
             }
   \date{Received / accepted}

   \titlerunning{BeppoSAX uncovers the hidden Seyfert 1 nucleus in the Seyfert 2 galaxy NGC 2110}

   \authorrunning{G. Malaguti et al.}

   \maketitle

   \begin{abstract}
The Seyfert 2 galaxy NGC 2110 has been observed with BeppoSAX
between 0.5 and 150 keV. The high energy instrument onboard, PDS, 
has succeeded in measuring for the first time the spectrum
of this source in the 13-150 keV range. 
The PDS spectrum, having a photon index $\Gamma\simeq1.86$ is
fully compatible with that expected from a Seyfert 1 nucleus.
In the framework of unified models,
the harder ($\Gamma\simeq1.67$) 2--10 keV spectrum is well 
explained assuming the presence of a complex partial + total
absorber ($N_{\rm H}\simeq30\times10^{22}$cm$^{-2}$$\times\sim$25\%
+$N_{\rm H}\sim4\times10^{22}$cm$^{-2}$$\times\sim$100\%).
The high column density of this complex
absorber is consistent both with the FeK$_\alpha$ line strength and with
the detection of an absorption edge
at $E\simeq7.1$ keV in the power-law spectrum.

   \end{abstract}

%

\section{Introduction}
Active Galactic Nuclei (AGN) classified as
Narrow Emission Line Galaxies (NELG) were first 
discovered in X-rays because of their intense 2--10 keV emission.
Like Seyfert 2 galaxies, their optical spectra are dominated 
by narrow emission lines. This fact, together with
the presence of a broad H$\alpha$ feature
in the spectrum of a few of them (Shuder 1980), has led to the
suggestion that NELG can form a transition class between Seyfert 1
and Seyfert 2 galaxies (e.g. Lawrence and Elvis 1982). 
The NELG NGC 2110 ($z$=0.0076) was first observed in X-rays with SAS-3
(Bradt et al. 1978), and then by HEAO 1 (Mushotzky 1982), 
and by EXOSAT (Turner \& Pounds 1989). 
The 2--10 keV ASCA data revealed a moderately flat ($\Gamma\sim1.69$) 
absorbed power law plus FeK$_\alpha$ line spectrum attenuated by partial 
covering material, while GINGA did not measure a significant reflection
component (Hayashi et al. 1996).
The scenario described by the data available prior to the present work
indicates (Smith \& Done 1996) the possibility that the 2--10 keV spectrum of
NGC 2110 is intrinsically flatter than the $< \Gamma > \sim 1.9$ slope
observed for Seyfert 1 galaxies by GINGA (Nandra \& Pounds 1994). 
This situation, if confirmed, would clearly pose questions to the
unified models (Antonucci 1993). 
In the following, we present BeppoSAX observation of NGC 2110 which highlights 
the key role played by the measurement of the spectrum 
above 20 keV in disentangling the intrinsic nuclear emission.

\section{Observation and data reduction}

The BeppoSAX X-ray observatory (Boella et al. 1997a) is a major programme
of the Italian Space Agency with participation of the Netherlands Agency for 
Aereospace Programs. 
This work concerns results obtained with three of the 
Narrow Field Instruments (NFI) onboard:
the Low Energy Concentrator Spectrometer (LECS; Parmar et al. 1997), 
the Medium Energy Concentrator Spectrometers 
(MECS; Boella et al. 1997b), and the Phoswich 
Detector System (PDS; Frontera et al. 1997).
LECS and MECS operate in the 0.1--4.5 keV and 1.5--10 keV spectral 
bands respectively, while 
PDS covers the 13--300
keV band. 

BeppoSAX NFI pointed at NGC 2110 from Oct 12th, to Oct 14th, 1997.
The effective exposure times were $4.16\times10^4$ s for the LECS,
$8.41\times10^4$ s for the MECS, and $3.30\times10^4$ s for the PDS.
Spectral data were extracted from region centred in NGC 2110
with a radius of 2$'$ and 4$'$ for the MECS and the LECS respectively. 
LECS and MECS background subtraction was performed by means of blank sky spectra
extracted from the same region of detector's field of view.
The net source count rate was $5.8\times10^{-2}$ c/s in the LECS
(0.5--4.5 keV), and 0.23 c/s in the two MECS (2--10 keV).
The detected count rates correspond, for the best-fit model 
(see section 3.3), to 
F$_{\rm 0.5-2\; keV}=5.9\times10^{-13}$,
F$_{\rm 2-10\; keV}=3.0\times10^{-11}$, and 
F$_{\rm 13-100\; keV}=6.1\times10^{-11}$
(in units of erg cm$^{-2}$ s$^{-1}$).

\section{Spectral analysis}

LECS and MECS data were rebinned in order to sample the energy resolution of the
detector with an accuracy proportional to the count rate: one channel
for LECS and 5 channels for MECS.
Spectral data from LECS (0.5--4.5 keV), MECS (2--10 keV), and PDS (13--150 keV) have 
been fitted simultaneously. Normalization constants have been introduced to allow
for known
differences in the absolute cross-calibration between the detectors. The values
of the two constants have been allowed to vary in a $\pm$10\% 
interval around the suggested (Cusumano et al. 1998, Dal Fiume 
private communication) values
(C$_{\rm LECS}$/C$_{\rm MECS}$=0.65, C$_{\rm PDS}$/C$_{\rm MECS}$=0.90).
The best fit values turned out to be within $\sim$5\% of the suggested ones.
The spectral analysis has been performed by means of the {\sc XSPEC 10.0} package,
and using the instrument response matrices released by the BeppoSAX Science Data 
Centre in September 1997. 
All the quoted errors correspond to 
90\% confidence intervals for one interesting parameter ($\Delta\chi^2$ of 2.71).
Source plus background light curves did not indicate significant flux variability.
Therefore the data from the whole observation were summed together for the 
spectral analysis.

\begin{table*}
 \centering
 \caption[line]{Spectral fits results. The Fe edge energy was fixed at 7.1 keV.}
 \begin{tabular}{cc c cc c cc c}\hline
$\Gamma_{2-10\;{\rm keV}}$ & $\Gamma_{13-150\;{\rm keV}}$ & $N_{\rm H}^{\rm Total}$          & 
$E$(FeK$_\alpha$)      & $EW$(FeK$_\alpha$)         & $\tau_{\rm Fe}$ &
$N_{\rm H}^{\rm Part.}$          & $C_{\rm F}$ & $\chi^2/\nu$ \\
                          &                             & $(\times10^{22}\;{\rm cm}^{-2})$ &
(keV)    & (eV)      &                                  & $(\times10^{23}\;{\rm cm}^{-2})$ &             & \\
 \hline 
$1.67^{+0.06}_{-0.06}$    & -                           & $3.7^{+0.3}_{-0.3}$
&
$6.41^{+0.07}_{-0.07}$ & $176^{+20}_{-39}$ & -                                & -
& -                    & 164.3/183$^\dagger$ \\
 \hline 
$1.66^{+0.06}_{-0.06}$             & -                           &
$3.4^{+0.4}_{-0.3}$ &
$6.41^{+0.07}_{-0.07}$        & $157^{+25}_{-29}$ & $0.13^{+0.07}_{-0.07}$
& - & -             & 159.5/182$^\dagger$ \\
 \hline 
$1.58^{+0.06}_{-0.05}$    & $1.97^{+0.07}_{-0.14}$      & $3.5^{+0.3}_{-0.5}$            &
$6.42^{+0.06}_{-0.07}$ &$158^{+28}_{-32}$& $0.13^{+0.06}_{-0.05}$    & -                                & -                    & 167.0/188 \\
 \hline 
\multicolumn{2}{c}{$1.82^{+0.14}_{-0.12}$}              & $4.1^{+0.5}_{-0.3}$            &
$6.41^{+0.07}_{-0.06}$ &$152^{+23}_{-22}$& -                         & $3.24^{+2.06}_{-2.98}$           &$0.25^{+0.16}_{-0.14}$& 171.1/188 \\
 \hline 
\multicolumn{9}{l}{$^\dagger$ PDS data are not included.}
 \end{tabular}
\end{table*}

All the models used in what follows contain an additional term to allow for the
absorption of X-rays due to our Galaxy that in the direction of NGC 2110 amounts
to $1.86\times10^{21}$ cm$^{-2}$ (Elvis et al. 1989). The energy values of the 
emission line(s) and absorption edge(s) are given in
the reference system of the emitting source, unless otherwise stated.

\subsection{The 0.5-10 keV spectrum}

The LECS-MECS spectrum has been fitted with a simple absorbed power law model
with the addition of a narrow gaussian line to account for FeK$_\alpha$ emission.
The fit is
satisfactory ($\chi^2=164.3/183$), and results in a flat spectrum with 
photon index $\Gamma=1.67\pm0.06$
and N$_{\rm H}=(3.7\pm0.3)\times10^{22}$ cm$^{-2}$. The line feature is centered
at 
$6.41\pm0.07$ keV, with an equivalent width (EW) of 176$^{+20}_{-39}$ eV. 
If the line width is allowed to vary,
the additional parameter gives only a negligible statistical improvement in the fit
($\chi^2=164.0/182$), and in any case, the upper limit obtained,
$\sigma<0.08$ keV, is consistent with the 
energy resolution of the MECS, which at 6.4 keV is $\sim$7\% 
(FWHM, Boella et al. 1997b).
The line width was then frozen to zero for the subsequent analysis.
The introduction of an absorption edge consistent with neutral Fe
($E_{\rm edge}\equiv7.1$ keV, $\tau=0.13\pm0.07$)
turns out in a significant improvement of the fit ($\chi^2=159.5/182$, corresponding to
$>$95\% confidence), and gives a first hint for the presence of an additional absorber. 
The LECS+MECS 2--10 keV data on NGC 2110 confirm basically the previous results obtained 
by GINGA (Hayashi et al. 1996), BBXRT (Weaver et al. 1995), and ASCA (Hayashi et al. 1996;
Turner et al. 1997), and allow a better measurement of the optical depth of the Fe edge
detected by ASCA.
A second hint for the presence of a complex absorber is given by the fact that
the observed line intensity is too high to be produced by transmission through
the measured absobing column (Leahy \& Creighton 1993).
The consistency is reached only if we add a second
absorber responsible for the observed FeK edge: assuming the Fe cross
section
given by Leahy \& Creighton (1993), the measured optical depth $\tau\simeq0.13$
corresponds to an equivalent hydrogen column density 
$N_{\rm H}=(1.1\pm0.6)\times10^{23}$ cm$^{-2}$,
which is then consistent with the measured Fe line EW.

\subsection{The Seyfert 1 nucleus in the PDS spectrum}

The most important result of the present work is that
the high energy spectrum (13--150 keV) is well fitted
($\chi^2=7.4/7$) by a simple power law model with
$\Gamma=1.86^{+0.14}_{-0.13}$. This is the first evidence for the presence of a steep, 
X-ray spectrum in NGC 2110, previously classified as a ``flat-spectrum''
source (Smith \& Done 1996). In the effort of verifying, and quantifying the significance of the
spectral steepening above 10 keV, a broken power law model was used. 
For simplicity the knee of the broken
power law was frozen at 10 keV, but a knee with free energy did not affect the results 
significantly.
The model results in a good ($\chi^2=167.0/188$) fit to the data and the two photon indices
are $\Gamma_{2-10}=1.58^{+0.06}_{-0.05}$ and  $\Gamma_{\rm 13-150}=1.97^{+0.07}_{-0.14}$.
   \begin{figure}
      \label{bkn1}
\psfig{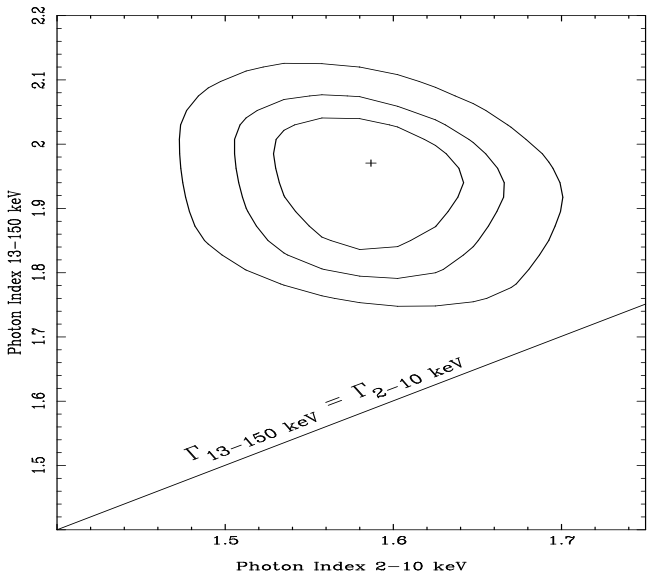}
      \caption[]{Confidence (68\%, 90\%, and 99\%) contour plot of the 2--10 keV
and 13--150 keV photon indices for a double power law model. The two indices are different
at $>$99\% confidence.}
   \end{figure}
Figure 2 shows clearly that the two indices are different at $>$99\% confidence, 
and that
$\Gamma_{\rm 13-150}$ is consistent with the canonical Seyfert 1 X-ray spectrum 
slope (Nandra \& Pounds 1994).
The observed $\Delta\Gamma\sim0.4$ remains significant also if we introduce
the systematics due to cross-calibration between MECS and PDS which 
are of the order of 3\%, or
$\Delta\Gamma\sim0.06$ (Cusumano et al. 1998).

\subsection{The broad-band 0.5-150 keV spectrum}

Given the 13-150 keV slope, the X-ray spectrum below 10 keV must be interpreted as the result 
of some kind of reprocessing of the primary emission. In what follows we have tried to model the
broad-band BeppoSAX spectrum of NGC 2110 in order to reconcile the observed 2-10 keV 
flatness with the steepening observed at $E>13$ keV.
As a first attempt we considered the hypothesis 
that the observed spectrum might include a reflection component
that would produce a flatter spectrum below 20--30 keV and a steepening 
at higher energies. The reflector(s), in the unified model scenario,
could be identified with the accretion disk, the Broad Line Region (BLR), 
or the circumnuclear torus.
The reflection component obtained with this model ({\sc pexrav} in XSPEC)
is low ($<0.17$ and $\sim0.5$ for $E_{\rm cutoff}$$\equiv$10000 and 50 keV
respectively) and is therefore not consistent with the Fe line EW
(which requires a reflection component $\sim1$). Moreover 
the spectral index remains flat ($\Gamma=1.69^{+0.05}_{-0.04}$). 
On the other hand the measured absorption column density cannot
explain the observed Fe line EW and FeK edge optical depth,
unless a large, $\sim10$, Fe overabundance is introduced in the system.

A further physical situation that can, in an AGN environment, 
harden and therefore conceal an intrinsic steep spectrum is the 
presence of a complex absorber. We applied this model and
the result (the spectrum and residuals are shown in figure 2)
was very satisfying ($\chi^2=171.1/188$), with the slope increased to 
$\Gamma=1.82^{+0.14}_{-0.12}$, while the additional partially covering
column density was 
$N_{\rm H}^{\rm pc}=(3.24^{+2.06}_{-2.98})\times10^{23}$ cm$^{-2}$, 
for a covering fraction $C_{\rm F}=0.25^{+0.16}_{-0.14}$.
Contours of $C_{\rm F}$ versus $N_{\rm H}^{\rm pc}$ are reported in 
figure \ref{cov1}.
The best fit value of $N_{\rm H}^{\rm pc}$ is consistent with the
optical depth inferred from the additional absorption edge (see section 3.1), 
and, as a confirmation of this result,
the addition of a FeK edge is not required by this model.
The partial covering absorbing column is therefore consistent both with the observed
depth of the FeK edge, and also with the measured Fe line EW.
   \begin{figure}
\psfig{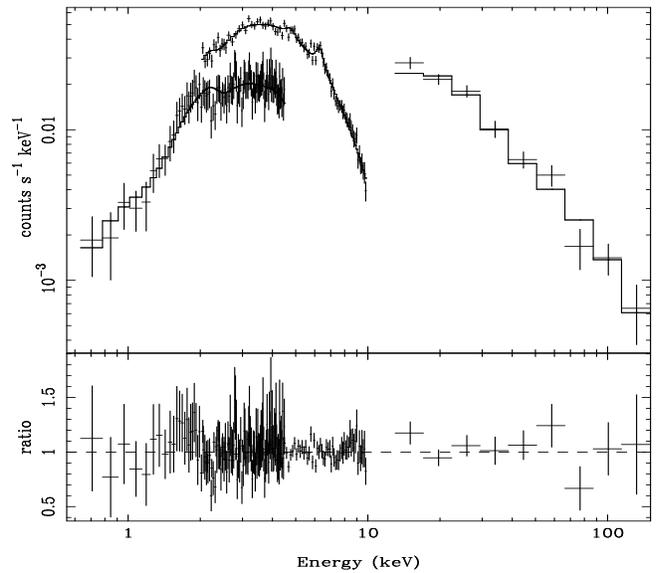}
      \caption[]{Broad band spectrum with residuals for the dual absorber
plus FeK line model (see text).} 
         \label{euf}
   \end{figure}

\section{Discussion and Conclusions}

BeppoSAX high energy instrument, PDS, has succeeded in measuring 
for the first time the X-ray spectrum
of the Seyfert 2 galaxy NGC 2110 at energies above 15 keV. The spectrum,
having a photon index $\Gamma=1.86^{+0.14}_{-0.13}$ 
is fully compatible with what expected from a Seyfert 1 nucleus (Nandra and
Pounds 1994). The interpretation of the flat ($\Gamma=1.67\pm0.06$) 
spectrum observed between 2 and 10 keV in terms of reflection is neither
required by the data ($R<0.5$) nor is compatible with them,
since the observed Fe EW would imply a reflection a factor
of 5 greater than the observed upper limit.
   \begin{figure}
\psfig{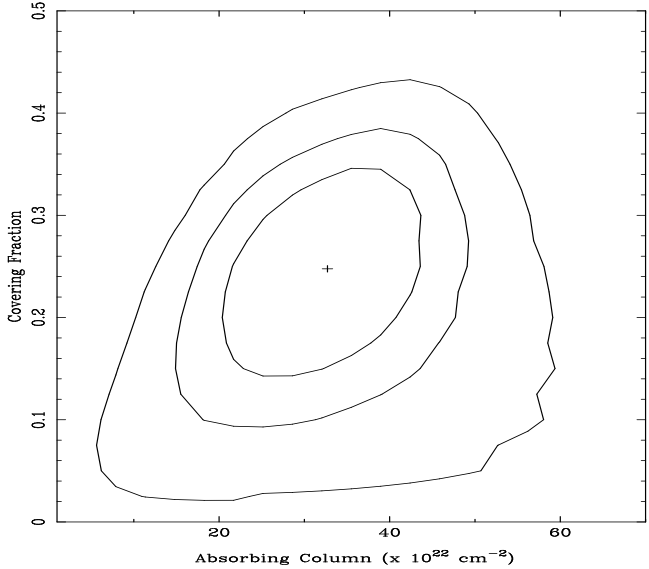}
      \caption[]{Confidence contours for the covering fraction vs equivalent Hydrogen
column density of the partially covering material in the dual absorber model.} 
         \label{cov1}
   \end{figure}
The flat 2--10 keV spectrum is reconciled with unified models and with the 
$E>13$ keV power-law slope assuming the presence of a complex absorber.
In the framework of unified models 
the two ($N_{\rm H}^{\rm Partial}$ and $N_{\rm H}^{\rm Total}$) 
column densities can be associated
with the torus or with a combination of the BLR plus an additional
absorber.

In the first scenario the dual absorber can be explained in terms
of a complex configuration of the torus: an inner, denser and 
tidally disrupted region plus an outer more homogeneous one
which account for the partial and total absorbing column respectively.
Alternatively, the complex absorber can be associated with the presence
of a torus ``atmosphere'' (Feldmeier et al. 1998).
In the second case the partial absorber is physically associated
with the clouds of the BLR, assuming that they are small if 
compared with the nuclear X-ray source size or the source appears
much larger due to nearby X-ray scattering induced by highly
ionized gas.
The upper limit
on the BLR clouds Doppler velocity ($<$ few 10$^{3}$ km/s)
deduced from the FeK width
is also consistent with this hypothesis. The second (total) absorber
can again be the torus itself (Hayashi 1996) or the Intermediate
Line Region introduced by Cassidy \& Raine (1997) for the Seyfert 1.5
NGC 4151.
In NGC 2110 no obvious broad line components have been detected
(Veilleux et al. 1997). Moreover the observed nuclear reddening 
provides only a lower limit ($A_V\ge4.6$ mag) on the amount of
obscuration to the nucleus (Mulchaey et al. 1994). 
Therefore the present data do not allow to firmly assess 
where the absorption regions are to be located and therefore
to discriminate among the proposed scenarios.

The interpretation of the flat 2--10 keV spectrum of Seyfert 2 galaxies as the 
result of an artifact caused by the presence of a complex absorber is not new, 
since it has been suggested to explain the X-ray spectrum of NGC 5252
(Cappi et al 1996), IRAS 04575--7537 (Vignali et al. 1998), and NGC 7172
(Guainazzi et al. 1998).

The detection of an ``excess'' FeK edge in the spectrum of ESO 103--G35, NGC 7314
(Turner et al. 1997), and possibly NGC 7582 (Xue et al. 1998), suggests that
an additional unmodelled absorber is present in these sources.
While this could indicate that the dual absorber scenario can be applicable to
Seyfert 2 galaxies and NELGs in general, other observational results show that this model 
is not the universal solution for reconciling flat and ``canonical'' (i.e. $\Gamma\sim1.9$)
sources. Namely, the BBXRT spectrum of NGC 4151 results in a flat 
($\Gamma\sim1.5$) power law also after the addition of a complex ionized absorber 
(Weaver et al. 1994). Moreover, Turner et al. (1997) 
show that the ASCA spectrum of NGC 2110 
remains flat ($\Gamma\sim1.5$) also after the addition of ionized material 
partially covering the source. Finally, the recent results on the ASCA spectrum 
on the Seyfert 1.5 LB 1727 shows an unattenuated flat ($\Gamma\sim1.6$) power law
(Turner et al. 1999).

Clearly, the easiest interpretation of the flat spectrum of NGC 2110 is that
the source is actually intrinsically flat. In this case
the observed spectral steepening at $E>13$ keV
could be the exponential cut-off of the primary power law. We have tested this model 
with our data. The best fit ($\chi^2=170/189$) gives a photon index 
$\Gamma=1.64^{+0.09}_{-0.13}$ with a cutoff energy value 
$E_{\rm C}=65^{+252}_{-36}$. We cannot, therefore, discriminate among
this model and the dual absorber one on a statistical basis.
As far as the physical interpretation in concerned, however, we
prefer the interpretation of the 0.5--150 keV spectrum of NGC 2110
in terms of a dual absorber for several reasons.
The exponentially cut-off power law fails in explaining the observed
FeK line EW and absorption edge optical depth, unless a large
($\sim 10$) iron overabundance is introduced.
Moreover, the high energy ($E>13$ keV) spectrum in the dual absorber model 
is fully consistent with the slope measured using the PDS data only.
Finally, the results listed above regarding flat X-ray spectrum Seyfert 
galaxies have all been obtained with instruments operating only up
to 10 keV, while we have shown in this work that the steep, intrinsic spectrum 
of NGC 2110 is measurable only above 13--20 keV.

In summary, the new result on the steepening of the NGC 2110 spectrum above 13 keV
is well explained assuming the presence of a complex absorber. 
On the other hand, this model fails in explaining 
the flat 2--10 keV spectrum observed in several other Seyfert galaxies, and cannot,
therefore be accepted as a universal explanation of ``flat-spectrum'' Seyfert galaxies.
It is important to point out that this issue can be fully addressed only with 
future broad-band observations of an extended sample of flat-spectrum Seyfert galaxies 
covering the entire 2--100 keV region.

\begin{acknowledgements}
This research has made use of SAXDAS linearized and cleaned event files produced at
the BeppoSAX Science Data Centre. G.M., G.G.C.P., M.C., A.C., and L.B. acknowledge 
financial support from the Italian Space Agency. 
We would like to thank the referee Dr. J. Turner for the very useful comments
which have significantly improved the quality of this work.
\end{acknowledgements}

\end{document}